\documentclass[final,3p,times]{elsarticle}

\usepackage{amssymb}
\usepackage{amsmath}
\usepackage{textgreek}
\usepackage{float}

\usepackage{lineno}

\journal{Journal of Subatomic Particles and Cosmology}

\begin{document}

\begin{frontmatter}


\title{Non-Monotonicity of $p_{\rm T}$ Correlations in Au + Au Collisions at RHIC }

\author[Rutik Manikandhan]{Rutik Manikandhan (for the STAR collaboration)}
\affiliation[aaa]{organization={University of Houston},
             addressline={4800 Calhoun Rd.},
             city={Houston},
             postcode={77204},
             state={TX},
             country={USA}}

\begin{abstract}
We report the first measurements of two-particle transverse momentum correlations for mid-rapidity charged particles in Au+Au collisions at $\sqrt{s_{\rm NN}} =$ 3.0, 3.2, 3.5, 3.9, 4.5, 5.2 and 7.7 GeV recorded by the STAR experiment. The results are compared with previous STAR measurements from the Beam Energy Scan Phase I (BES-I) and with transport model calculations. The measured two-particle $p_{\rm T}$ correlators exhibit an approximate power-law scaling with the number of participating nucleons ($N_{\rm part}$), consistent with expectations from an independent-source scenario. This scaling is least well described at the lowest collision energy, $\sqrt{s_{\rm NN}} = 3.0$~GeV. Furthermore, a non-monotonic energy dependence of the $p_{\rm T}$ correlations is observed in central collisions, representing the first systematic observation of such behavior as a function of collision energy.
\end{abstract}


\end{frontmatter}



\section{Introduction}
\label{intro}

The Beam Energy Scan (BES) program at RHIC probes the QCD phase structure over a broad range of baryon densities~\cite{Chen:2024aom}. The STAR fixed-target (FXT) program extends this reach to very high baryon chemical potential, up to $\mu_{B}\approx 760$~MeV~\cite{STAR:2020dav}, overlapping the region of interest for the conjectured QCD critical point~\cite{PhysRevD.101.054032}. In this work we present two-particle transverse-momentum correlation measurements in Au+Au collisions at $\sqrt{s_{\rm NN}}=3.0$--$7.7$~GeV, which are sensitive to dynamical fluctuations beyond purely statistical baselines~\cite{STAR:2019dow,ALICE:2014gvd}. The preliminary results shown in the presentation have since been finalized and published in~\cite{2xsn-rgx3}.

\section{Experimental setup and analysis}
\label{setup}
Transverse momentum correlations are characterized by the two-particle correlation function
defined using the covariance:

\begin{equation} \label{eq3}
\langle \Delta p_{{\rm T},i} \Delta p_{{\rm T},j} \rangle
= \left\langle \frac{\sum\limits_{i,j,\,i\neq j}^{N_{\rm ch}} (p_{{\rm T},i} - \langle\langle p_{\rm T} \rangle\rangle)(p_{{\rm T},j} - \langle\langle p_{\rm T} \rangle\rangle)}{N_{\rm ch}(N_{\rm ch} - 1)} \right\rangle_{\rm ev} ,
\end{equation}

which may be evaluated in the equivalent $Q$-vector form

\begin{equation} \label{eq3b}
\langle \Delta p_{{\rm T},i} \Delta p_{{\rm T},j} \rangle
= \left\langle \frac{Q_1^2 - Q_2}{N_{\rm ch}(N_{\rm ch} - 1)} \right\rangle_{\rm ev}
- \left\langle \frac{Q_1}{N_{\rm ch}} \right\rangle_{\rm ev}^{2} ,
\end{equation}

where $Q_n = \sum_{i=1}^{N_{\rm ch}} (p_{{\rm T},i})^n$. The $Q$-vector form in Eq.~\ref{eq3b} subtracts self-pairs via $Q_1^2-Q_2$ and can be evaluated with a single particle loop per event~\cite{PhysRevC.105.014906}. Here, $\langle\cdots\rangle_{\rm ev}$ denotes an event-ensemble average in which each event is weighted equally, and $\langle\langle p_{\rm T} \rangle\rangle = \langle Q_1/N_{\rm ch}\rangle_{\rm ev}$; the equivalence of Eqs.~\ref{eq3} and~\ref{eq3b} follows from this convention.

We report the relative dynamical correlation
\begin{equation}
C_{p_{\rm T}} \equiv
\frac{\sqrt{\langle \Delta p_{{\rm T},i}\Delta p_{{\rm T},j} \rangle}}
{\langle\langle p_{\rm T}\rangle\rangle}.
\end{equation}

\section{Results and Discussion}
\label{results}

\begin{figure}[!t]
\centering
\includegraphics[width=0.68\linewidth]{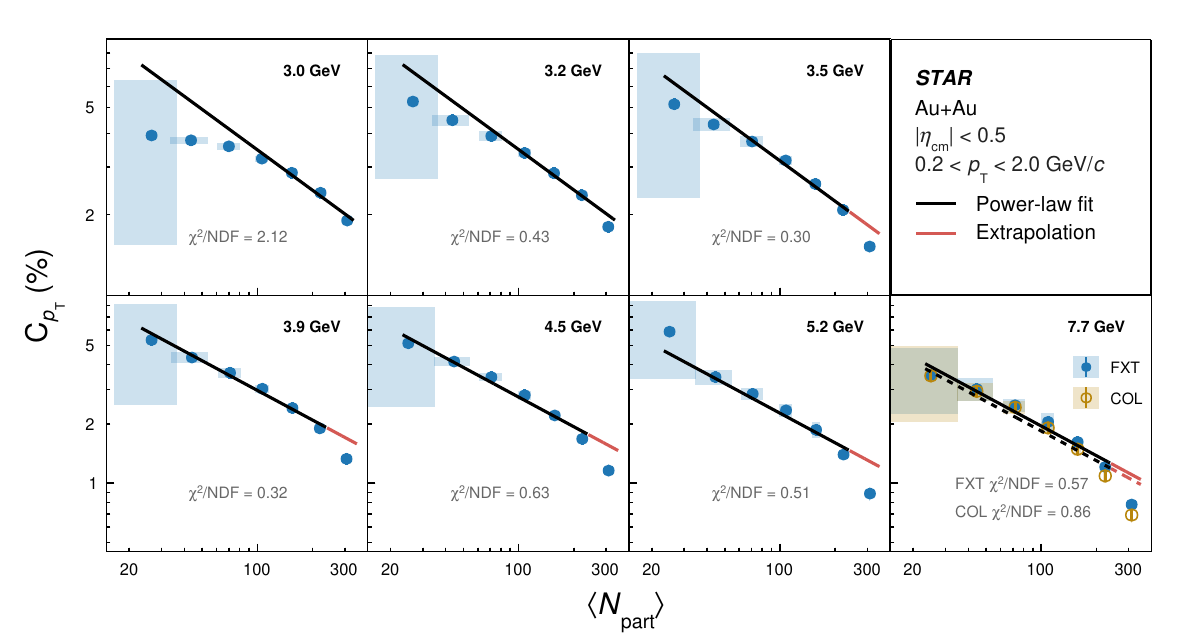}
\caption{\label{fig:centrality}The relative dynamical transverse momentum correlation \(C_{p_{\rm T}}\) as a function of the number of participating nucleons \(\langle N_{\rm part}\rangle\) in Au+Au collisions at several collision energies. The 7.7~GeV data include both Fixed-Target (FXT, blue circles) and Collider (COL, yellow circles) configurations. Vertical bars are statistical uncertainties, shaded boxes systematic. Black lines are power-law fits with the exponent fixed to $-0.5$, performed over all 10\%-wide centrality bins beyond 10\%; red lines extrapolate these fits into the 0--10\% bin, which is excluded from the fit. Charged particles are selected within \(0.2 < p_{\rm T} < 2.0\)~GeV/$c$ and \(|\eta_{\rm cm}| < 0.5\).}
\end{figure}

Figure~\ref{fig:centrality} shows $C_{p_{\rm T}}$ as a function of centrality for several collision energies; the correlations decrease towards central collisions. In the absence of collective or critical effects they are expected to follow a statistical baseline arising from the superposition of many independent particle-emitting sources, giving an approximate power-law dependence $A(\sqrt{s_{\rm NN}})/\sqrt{N_{\rm part}}$ that reflects the dilution of fluctuations with increasing system size. Deviations from this scaling therefore probe non-trivial dynamics, including modifications to the effective degrees of freedom or density-dependent effects, which may be enhanced in central collisions and could reflect proximity to a critical point~\cite{PhysRevC.92.024915}.

Because the available multiplicity at these energies does not permit finer centrality granularity, and because a fixed bin width is required for a consistent centrality-dependence study, we use 10\%-wide bins throughout; 5\%-wide bins cannot be reliably defined in peripheral collisions at the lowest energies. We follow a constrained procedure in which the power-law exponent is fixed to the expected baseline value of $-0.5$, leaving the normalization $A(\sqrt{s_{\rm NN}})$ as the only free parameter; at LHC energies the exponent can instead be treated as free~\cite{ALICE:2014gvd,ATLAS:2024jvf}, owing to the higher multiplicity and better centrality resolution. The fit is performed over all centrality bins beyond 10\%, and the resulting curve is extrapolated into the 0--10\% bin, which is excluded from the fit and subsequently compared with that extrapolation. Statistical and systematic uncertainties are added in quadrature in the fit, with systematic uncertainties treated as uncorrelated between centrality bins; since a substantial fraction of the systematic uncertainty is common to all bins, this treatment is conservative for both the goodness-of-fit and the extrapolated deviation.

At the lowest energy, $\sqrt{s_{\rm NN}} = 3.0$~GeV, where the system is expected to be dominated by hadronic interactions throughout its evolution, the constrained power-law fit describes the data noticeably less well ($\chi^{2}/{\rm NDF} = 2.12$). While this does not by itself constitute a rejection of the independent-source picture, it is consistent with the expectation that the scaling assumption becomes least applicable at the lowest beam energy. At the higher energies, and even under this conservative treatment, we observe a clear and increasing deviation of the 0--10\% bin from the extrapolated baseline as $\sqrt{s_{\rm NN}}$ rises from 3.2 to 7.7~GeV, suggesting additional dynamical contributions beyond trivial statistical fluctuations.

\begin{figure}[!t]
\centering
\includegraphics[width=0.52\linewidth]{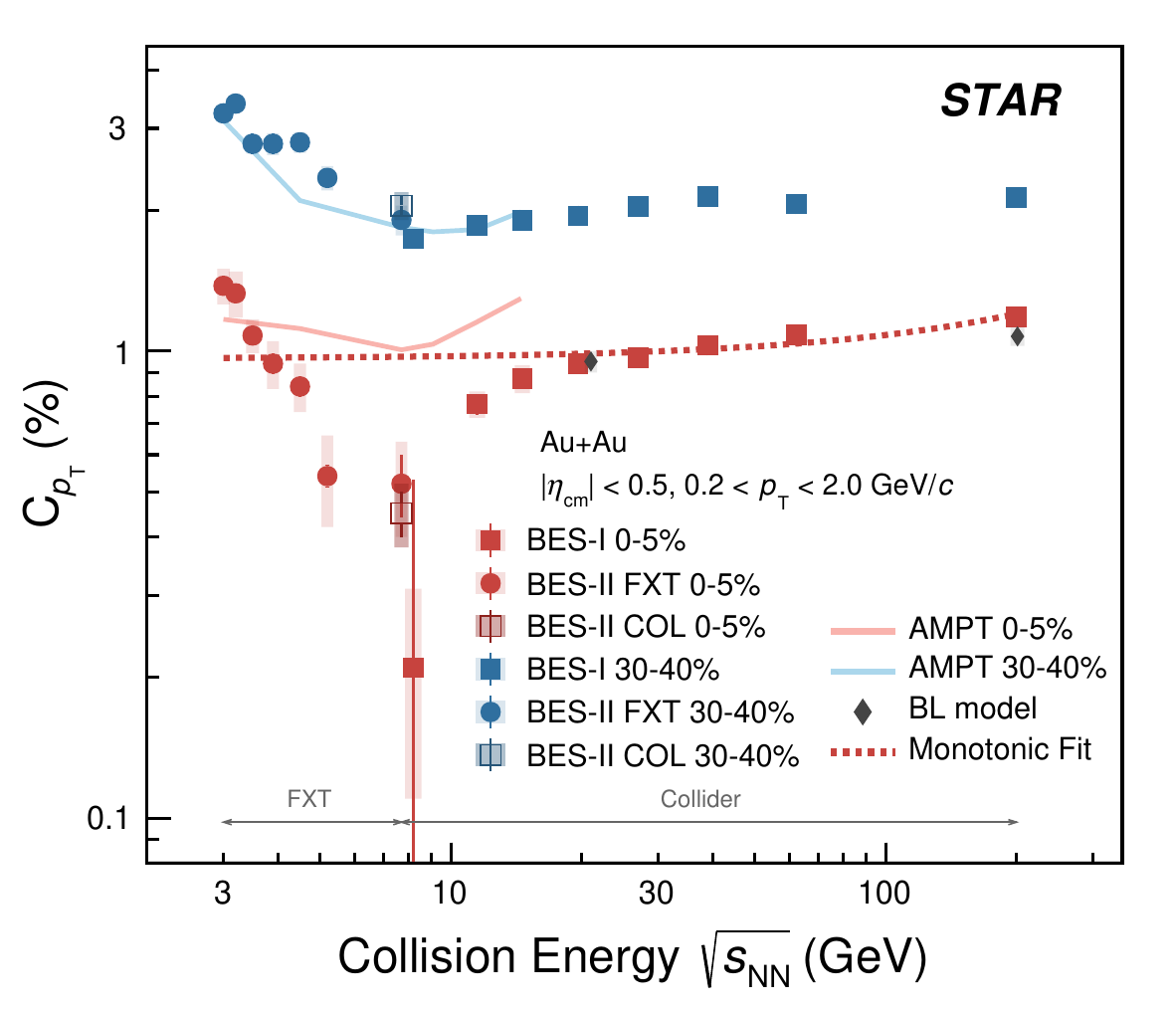}
\caption{\label{fig:Energy}The relative dynamical transverse momentum correlation \(C_{p_{\rm T}}\) as a function of collision energy \(\sqrt{s_{\rm NN}}\) in Au+Au collisions. Results are shown for 0--5\% (red) and 30--40\% (blue) centralities from BES-I, BES-II Fixed-Target (FXT) and BES-II Collider (COL) datasets; the new BES-II collider point at 7.7~GeV is a hollow square and the BES-I 7.7~GeV point is shifted horizontally for visibility. Vertical bars are statistical uncertainties, shaded boxes systematic. Charged particles are selected within $0.2 < p_{\rm T} < 2.0$~GeV/$c$ and $|\eta_{\rm cm}| < 0.5$. AMPT~\cite{PhysRevC.111.024911} (solid lines) is shown for both centralities and Boltzmann--Langevin~\cite{PhysRevC.95.064901} predictions as black diamonds. The dotted red line is the monotonic reference fitted to the 0--5\% data.}
\end{figure}

Figure~\ref{fig:Energy} shows the relative dynamical correlation $C_{p_{\rm T}}$ as a function of $\sqrt{s_{\rm NN}}$ for the most central bin (0--5\%) along with previous STAR measurements from BES~I~\cite{STAR:2019dow}, allowing a direct comparison over a wide range of beam energies. These data are compared against calculations from the A Multi-Phase Transport (AMPT)~\cite{PhysRevC.111.024911} and Boltzmann--Langevin (BL)~\cite{PhysRevC.95.064901} models.

Although AMPT describes the bulk $p_{\rm T}$ spectra reasonably well across these energies, it does not reproduce the measured correlations in detail~\cite{PhysRevC.111.024911}: in 0--5\% collisions it qualitatively follows the overall energy dependence but not the non-monotonic structure at intermediate energies, whereas the 30--40\% class is comparatively well described in both magnitude and its weaker energy dependence. Boltzmann--Langevin calculations~\cite{PhysRevC.95.064901}, which incorporate fluctuation--dissipation dynamics, show minimal energy dependence and are consistent with the STAR measurements at $\sqrt{s_{\rm NN}} = 19.6$ and 200~GeV. Several non-critical effects may also contribute to two-particle $p_{\rm T}$ correlations in this range, including baryon stopping, hadronic rescattering, resonance decays, and conservation laws; disentangling these from possible critical contributions will require dedicated model comparisons.

The new high-statistics data at $\sqrt{s_{\rm NN}} = 7.7$~GeV significantly reduce uncertainties compared to BES~I and remain fully consistent with previous measurements. To quantify deviations from a smooth energy dependence, the 0--5\% excitation function is compared with a strictly monotonic reference, taken as a first-order polynomial in $\sqrt{s_{\rm NN}}$ fitted over $19 < \sqrt{s_{\rm NN}} < 200$~GeV, where the measurements carry the smallest uncertainties. Statistical and systematic uncertainties are added in quadrature and treated as uncorrelated between energies. The significance is obtained from the $\chi^{2}$ of all measured 0--5\% points with respect to this reference (${\rm NDF} = 12$), converted to a two-sided Gaussian significance; the quoted value is conditional on the assumed functional form of the monotonic reference.

Using this procedure, a significance of approximately $5\sigma$ is obtained for the STAR 0--5\% central data. In contrast, the AMPT model calculations, analyzed under identical conditions and using the same methodology and uncertainty as the data, exhibit only a $\sim 1.4\sigma$ deviation, which is insufficient to support even weak evidence for non-monotonic behavior. This indicates that the procedure itself does not generate spurious significance in the absence of critical dynamics.

Applying the same analysis technique to mid-central collisions, a deviation of $\sim 2\sigma$ is observed in the 30--40\% centrality class. While this level of significance does not constitute conclusive evidence for non-monotonic behavior, it is consistent with the trend observed in the most central collisions and suggests a possible centrality dependence of the effect. This data-driven approach provides a robust and uniform framework for comparisons across centrality and is expected to motivate further theoretical and model investigations.

\section{Summary}
We have measured the relative dynamical correlation $C_{p_{\rm T}}$ in Au+Au collisions at $\sqrt{s_{\rm NN}} = 3$--$7.7$~GeV, corresponding to $\mu_B \approx 760$--$400$~MeV. The data exhibit an approximate power-law scaling with $\langle N_{\rm part}\rangle$ at most collision energies, consistent with independent-source expectations; this scaling is least well described at the lowest beam energy, where the system is expected to be hadronic throughout.

In central collisions the correlations show a non-monotonic dependence on collision energy with a significance of approximately $5\sigma$ relative to the monotonic baseline described above, and the measurements remain consistent between fixed-target and collider configurations. Because $C_{p_{\rm T}}$ is constructed from intensive quantities, the leading contribution from volume fluctuations cancels and the observable is correspondingly less sensitive to centrality bin-width effects than extensive cumulants~\cite{Luo:2014rea}; residual sensitivity to centrality resolution and to correlations between the centrality estimator and the analyzed particles is evaluated as part of the systematic uncertainty.

The energy region where the most pronounced dip is observed overlaps the range $\sqrt{s_{\rm NN}} = 4$--$6$~GeV quoted in Ref.~\cite{shah2024locatingqcdcriticalpoint} for a critical point located from contours of constant entropy density; that beam-energy range is obtained from the $\mu_B$ dependence of the chemical freeze-out curve and carries the associated model dependence. These results indicate non-trivial dynamical correlations in strongly interacting matter and provide new experimental constraints for models incorporating both critical and non-critical dynamics at high baryon density. Further theoretical work incorporating critical dynamics, and measurements of higher-order $p_{\rm T}$ cumulants, will be needed for a full interpretation.

\section*{Acknowledgements}
We thank the RHIC Operations Group and the RCF at BNL for their support. This work was supported in part by the National Key Research and Development Program of China (Contract Nos.~2024YFA1612600 and 2022YFA1604900) and by the U.S. Department of Energy, Office of Science.

\bibliographystyle{elsarticle-num}
\bibliography{sqm2026_template}

\end{document}